\documentclass[aps,prb,twocolumn,superscriptaddress,longbibliography]{revtex4-2}
\usepackage[colorlinks=true,linkcolor=blue,citecolor=blue]{hyperref}
\usepackage{amsmath,amssymb}
\usepackage[utf8]{inputenc}
\usepackage[T1]{fontenc}
\usepackage{url}
\usepackage{adjustbox}
\usepackage{subcaption}
\usepackage{graphicx}
\usepackage{bm}
\usepackage{bbm}
\usepackage{xcolor}
\usepackage{color}
\usepackage{ulem}
\usepackage{subfloat}
\usepackage{ragged2e}
\usepackage{caption}
\usepackage{braket}
\DeclareCaptionFormat{myformat}{\justifying#1#2#3}
\newsavebox{\measurebox}
\begin{document}


\title{Investigating topological in-gap states in non-Hermitian quasicrystal with unconventional $p$-wave pairing} 

\author{Shaina Gandhi}
\email{p20200058@pilani.bits-pilani.ac.in}
\affiliation{Department of Physics, Birla Institute of Technology and Science, Pilani 333031, India} 

\author{Jayendra N. Bandyopadhyay}
\email{jnbandyo@gmail.com}
\affiliation{Department of Physics, Birla Institute of Technology and Science, Pilani 333031, India}

\begin{abstract}
The interplay of onsite quasiperiodic potential, superconductivity, and non-Hermiticity is explored in a non-Hermitian unconventional superconducting quasicrystal described by Aubry-Andr\'e-Harper (NHAAH) model with $p$-wave pairing. In previous studies, the non-Hermiticity was only considered at the onsite quasiperiodic potential of the NHAAH model, and Majorana zero modes (MZMs) were observed under open boundary conditions (OBC) in this model. In this work, we study an NHAAH model with $p$-wave pairing, where non-Hermiticity is considered onsite by introducing complex quasiperiodic potential and asymmetry at the hopping part. Our analysis uncovers triple-phase transitions, where topological, metal-insulator, and unconventional real-to-complex transitions coincide at weak $p$-wave pairing strength. Additionally, instead of the MZMs observed in the symmetric hopping case, we observe the emergence of in-gap states under OBC in this model. These in-gap states are robust against disorder, underscoring their topological protection. Therefore, unlike the MZMs, which are very challenging to experimentally realize, these in-gap states can be used in topological quantum computational protocols.
\end{abstract}

\maketitle

\section{Introduction}
\label{sec1}

In disordered solids, all electronic states tend to be localized, while for solids with periodic lattice structures, all electronic states are delocalized as the strength of periodic modulation increases. However, incommensurate crystalline materials or quasicrystals form a class of solids that is intermediate to the crystalline and disordered class of solids. These solids show metal-insulator (MI) transition in one dimension, in which the system shifts from having delocalized (metallic) states to localized (insulating) states as a function of quasiperiodic modulation potential \cite{PhysRevB.23.6422}. The localization properties of these quasicrystals are extensively studied using the Aubry-Andr\'e-Harper (AAH) model \cite{AubryAndre, Harper}. This tight binding model comprises nearest-neighbor hopping in the presence of onsite cosine potential with irrational frequency relative to the lattice spacing. The AAH model exhibits a sharp localization transition due to its self-dual nature \cite{AubryAndre, PhysRevLett.48.1043, PhysRevLett.51.1198}. Recent developments in non-Hermitian systems have aroused substantial interest in investigating non-Hermiticity and disorder \cite{PhysRevLett.121.086803, PhysRevLett.121.026808, PhysRevLett.123.066404, padhi2024anomalousspectrumnonhermitianquasiperiodic}. The non-Hermitian systems give unique insights into localization phenomena such as non-Hermitian skin effect \cite{PhysRevLett.124.086801}, localization-delocalization transition induced by the presence of the random potential \cite{PhysRevLett.77.570, PhysRevB.56.8651}, etc. Recently, various extensions of the non-Hermitian AAH (NHAAH) model have received extensive attention \cite{yuce2018pt, PhysRevB.103.014201, PhysRevB.103.014203, longhi2019metal, jiang2019interplay, PhysRevLett.122.237601, PhysRevB.108.014204, PhysRevA.103.L011302, PhysRevB.104.024201}. Studies have shown that the NHAAH model with parity-time $(\mathcal{PT})$ symmetry demonstrates a triple phase transition, with simultaneous MI, topological, and $\mathcal{PT}$ transitions \cite{PhysRevLett.122.237601,PhysRevB.108.014204}. 

In Ref. \cite{PhysRevB.104.024201}, a different NHAAH model was introduced, incorporating non-Hermiticity in both quasiperiodic onsite potential and asymmetric hopping. This combination of non-Hermiticity and disorder reveals a new physical property. In contrast to prior models, this variant is not $\mathcal{PT}$ symmetric for any values of the system parameters. The existence of asymmetric hopping in this model leads to a non-$\mathcal{PT}$ symmetric nature, preventing the observation of a triple-phase transition. This analysis naturally leads to investigating whether a triple phase transition is still possible when non-Hermiticity is applied to the onsite potential, and the hopping remains asymmetric.

On a different front, unconventional superconductors have garnered substantial attention for their unique pairing mechanisms and ability to host exotic quantum states. Unlike conventional superconductors governed by $s$-wave pairing, the unconventional superconducting systems often exhibit $p-$ or $d-$ wave pairing that gives rise to non-trivial topological properties. These systems can host Majorana fermions, emergent quasiparticles with significant promise for fault-tolerant quantum computation \cite{Sarma2015}. Topological superconducting phases, characterized by bound Majorana zero modes (MZMs), have been extensively studied in Hermitian and non-Hermitian systems. A prominent example is the Kitaev chain, a one-dimensional $p$-wave superconducting model that serves as a paradigmatic framework for understanding the emergence of MZMs \cite{A_Yu_Kitaev_2001, Xiang_Ping_Jiang, Cheng_Gao,PhysRevB.106.094203, PhysRevB.94.125121,Vodola_2016,PhysRevB.93.104504,10.1093/ptep/ptad043,PhysRevB.94.125408,PhysRevB.110.094203,srinidhi2024quasimajoranamodespwavekitaev}. In recent years, there has been growing interest in studying the AAH model with $p$-wave pairing, uncovering interplay between disorder and superconducting pairing \cite{PhysRevB.93.104504,PhysRevResearch.3.013148,PhysRevLett.110.146404,PhysRevLett.110.176403,PhysRevB.93.104504,PhysRevB.94.125408,10.1093/ptep/ptad043,PhysRevB.107.014202}. Extensions of this work to the NHAAH model with $p$-wave pairing have revealed unconventional real-to-complex transitions in $\mathcal{PT}$-broken systems, where the real-valued superconducting pairing strength explicitly breaks $\mathcal{PT}$-symmetry \cite{PhysRevB.105.024514,PhysRevB.103.104203,PhysRevB.103.214202,PhysRevB.110.094203}. The non-Hermitian Kitaev chain with complex superconducting $p$-wave pairing exhibits $\mathcal{PT}$-symmetry without any correlation between topological and spontaneous $\mathcal{PT}$-symmetry breaking transitions  \cite{PhysRevA.92.012116}. Furthermore, the non-Hermitian Kitaev chain with nearest-neighbor pairing hosts MZMs, whereas the same model with long-range pairing demonstrates massive Dirac modes in the system \cite{PhysRevB.110.094203}. In addition, these systems lose the robustness of MZMs with increasing non-Hermiticity.

Motivated by these ongoing investigations, we aim to explore the interplay of $p$-wave pairing and asymmetric hopping in the NHAAH model with non-Hermiticity incorporated in both the onsite potential and the hopping terms. Our primary objective is to determine whether a triple-phase transition can occur within any parameter regime of this model. Additionally, we seek to examine the impact of non-Hermiticity in the hopping term on the existence and robustness of the topological properties of the system.

This paper is organized as follows: Section \ref{sec2} introduces the model and discusses its general properties. Section \ref{sec3} presents the phase diagram, demonstrating that a triple phase transition is achievable for weak pairing strength. Section \ref{sec4} shows that asymmetric hopping eliminates MZMs and gives rise to in-gap states. Section \ref{sec5} examines the robustness of these in-gap states against disorder.

\section{Model Hamiltonian} 
\label{sec2}

We consider the following tight-binding Hamiltonian that describes the NHAAH model with $p$-wave pairing:
\begin{equation}
	\begin{split}
H (h_1, h_2) &= \sum_{n=1}^L V \cos(2\pi\beta n + \theta + i h_1) c_n^{\dagger} c_n\\ &- t \sum_{n=1}^{L-1} (e^{-h_2} c_n^{\dagger} c_{n+1} +  e^{h_2} c_{n+1}^{\dagger} c_{n})\\ &+  \Delta \sum_{n=1}^{L-1} (\hat{c}_{n+1}^\dagger \hat{c}_n^\dagger + \hat{c}_{n} \hat{c}_{n+1}),
\label{Hamiltonian}
	\end{split}
\end{equation}
where $c^\dagger_n$ ($c_n$) represents the creation (annihilation) operator at site $n$, and $L$ is the total number of sites. The first term in the Hamiltonian describes the on-site potential, modulated by a cosine function of strength $V$ with spatial periodicity $\frac{1}{\beta}$. For irrational values of $\beta$, the system becomes quasiperiodic. A common choice of $\beta$ is the inverse of the golden mean ratio, i.e., $\beta = \frac{\sqrt{5}-1}{2}$. This irrational number is approximated by rational numbers $\beta = \frac{F_{n-1}}{F_{n}}$, where $F_n$ is the $n$-th Fibonacci number. The parameter $\theta$ is the phase factor that shifts the potential, and $h_1$ is a non-Hermitian parameter. The second term represents asymmetric hopping, exponentially biased in one direction. The parameter $h_2$ represents the other non-Hermitian parameter that controls the degree of asymmetry. The third term represents superconducting $p$-wave pairing, where $\Delta \in \mathbb{R}$ determines the pairing strength. We set the parameters $t$ and $V$ to {\it unity} to normalize hopping and potential energy scales, respectively. In this study, we set the number of sites $L=377$ and the parameter $\theta = 0$ (except for the case of topological transition). This model obeys time-reversal symmetry (TRS), defined as $T^{-1}_{+} H T_{+} = H$, where $T_{+}T_{+}^* = 1$, and $T_{+}$ corresponds to complex conjugation. Consequently, this model also satisfies a variant of particle-hole symmetry ($PHS^\dagger$), defined as $T^{-1}_{-}(iH)^*T_{-} = -iH$, where $T_{-}=PC $. Here, $\mathcal{P}$ denotes the parity (spatial reflection) operator, which acts as:
\[\mathcal{P}^{-1} c_n \mathcal{P} = c_{N+1-n},\]
and $\mathcal{C}$ is the charge conjugation operator defined as:
\[\mathcal{C} c_n \mathcal{C}^{-1} = i c_n^\dagger, \quad \mathcal{C} i \mathcal{C}^{-1} = -i.\]
Thus, this model is categorized in $AI$ or $D^\dagger$ symmetry class of non-Hermitian $38$-fold symmetry classification \cite{PhysRevX.9.041015}. 

In the absence of pairing, i.e., $\Delta = 0$, this model can be simplified to the model presented in Ref. \cite{PhysRevB.104.024201}, where the asymmetric hopping breaks the $\mathcal{PT}$ symmetry. Double-phase transitions are observed for this case, i.e., the MI and topological transitions coincide. Moreover, no real eigenvalues were observed due to asymmetric hopping. For $h_2 = 0$ and $\Delta \neq 0$, this model reduces to an NHAAH model with pairing and symmetric hopping. This model provides a testbed for a comprehensive analysis of the interplay of localization and topological phase transitions in non-Hermitian systems with $p$-wave pairing \cite{PhysRevB.103.214202}. The inclusion of the pairing term in the model introduces MZMs for various parameter regimes, and eigenvalues also become real-valued for the same parameter regime.

We include the pairing interactions in the Hamiltonian, given in Eq. \eqref{Hamiltonian}, by representing it in the Bogoliubov-de Gennes (BdG) basis
$\chi = (c_{0}, c_{0}^{\dagger}, \ldots, c_{N-1}, c_{N-1}^{\dagger})^{T}$, as follows:
\begin{equation}
H = \chi^{\dagger} H^{\rm BdG}\chi,
\end{equation}  
where
\begin{equation}
H^{\rm BdG} = 
\begin{pmatrix}
A_0 & B & 0 & \cdots & 0 & C \\
B^{\dagger} & A_1 & B & \cdots & 0 & 0 \\
0 & B^{\dagger} & A_2 & \cdots & 0 & 0 \\
\vdots & \vdots & \vdots & \ddots & \vdots & \vdots \\
0 & 0 & 0 & \cdots & A_{N-2} & B \\
C^{\dagger} & 0 & 0 & \cdots & B^{\dagger} & A_{N-1}
\end{pmatrix}
\label{BdG_Ham}
\end{equation}
In this representation, the Hamiltonian kernel $H^{\rm BdG}$ becomes a $2L \times 2L$ matrix with $A_n = V \cos(2\pi\alpha n + \theta_1 + i h_1) \sigma_{z}$ and the matrix $B$ is,  
\begin{equation}
B = \begin{bmatrix}
-t e^{-h_2} & -\Delta \\
\Delta & t e^{h_2}.
\end{bmatrix}.
\label{B_matrix}
\end{equation}
The matrix $C$ is represented under open boundary conditions (OBC) as
\begin{equation}
C = C^{\dagger}=\begin{bmatrix}
0 & 0 \\
0 & 0
\end{bmatrix}.
\label{C_matrix}
\end{equation}
On the other hand, for the periodic boundary conditions (PBC), the matrix $C$ is defined as $C = B^{\dagger}$. We aim to explore the possibility of triple-phase transition in this model with non-Hermiticity in hopping and potential terms.

\section{Phase Transitions} 
\label{sec3}

\begin{figure*}     
\begin{tabular}{ccc}
    \includegraphics[width=0.34\linewidth]{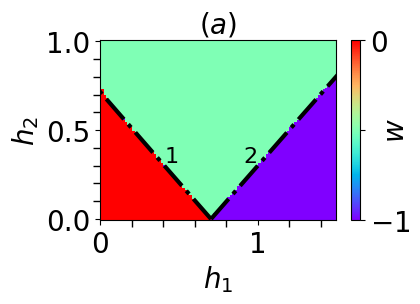}&
    \includegraphics[width=0.29\linewidth]{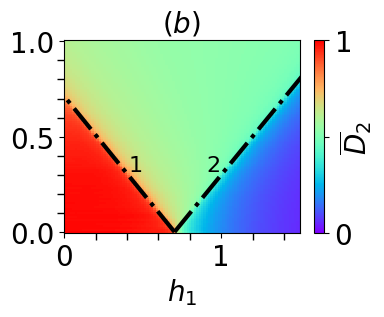}&
    \includegraphics[width=0.31\linewidth]{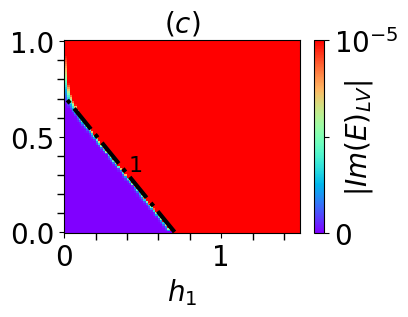}
\end{tabular}
\caption{The phase diagrams are presented for topological, MI, and unconventional real to complex transitions as a function of non-Hermitian parameters $h_1$ and $h_2$, for the pairing strength $\Delta = 0.01$. (a) The topological phases are identified using the winding number $w$. The red color represents the topological trivial region with $w = 0$. The cyan and purple colors correspond to the topological non-trivial region with winding numbers $-0.5$ and $-1$, respectively. The black dashed lines, marked by $1$ and $2$, represent the phase transition boundaries. (b) Averaged fractal dimensions are shown for the MI transition. The red, cyan, and purple colors correspond to delocalized, critical, and localized regions. Here again, the black dashed lines represent the phase boundaries. (c) The largest value of the imaginary parts of energy eigenvalues is presented to observe the unconventional real-to-complex transition. The purple color corresponds to the real region, while the red represents the complex region. The black dashed line demarcates the phase boundary between the two regions.}
\label{Phase} 
\end{figure*}

We study the variation of the transition point with the non-Hermitian parameters $h_1$ and $h_2$. The parameters $h_1$ and $h_2$ are tuned to identify whether triple phase transition is possible for a small pairing strength $\Delta = 0.01$. Here, the Hamiltonian $H^{\rm BdG}$ with the PBC is considered for studying all three phase transitions. The results are shown in Fig. \ref{Phase}. 
In Fig. \ref{Phase} (a), the topological phase transition is studied by calculating the winding number from the following relation:
\begin{widetext}
\begin{equation}
w(h_1, h_2) = \lim_{L \to \infty}\frac{1}{2\pi i} \int_{0}^{2\pi} d\theta \frac{\partial}{\partial \theta}\ln \left[\det {H^{\rm BdG}\left(\frac{\theta}{2L},h_1, h_2\right)- E_B}\right].
\label{wn}
\end{equation}
\end{widetext}
The winding number $w(h_1,h_2)$ counts the number of times the complex spectral trajectory encircles the base energy $E_B$ when the real phase $\theta$ varies from $0$ to $2\pi$. We have chosen $E_B = 0$ as the base energy for our analysis. The winding number values are determined by the strength of the non-Hermitian parameters. The phase diagram shows three distinct regions: $w(h_1, h_2) = 0, \,0.5$ and $1$, respectively. The fractional winding number in the critical region has been observed in various non-Hermitian systems \cite{PhysRevLett.116.133903, PhysRevA.97.052115,PhysRevB.103.214202}. The winding number zero corresponds to the topological trivial region, while the non-zero winding number corresponds to the topological non-trivial region. The black dashed line, marked by $1$, is the boundary between topologically trivial and non-trivial phases. This black dashed line is represented by the relation \[h_1 = \frac{\ln\left(2|-t\exp(-h_2)-\Delta|\right)}{V}.\] The other black dashed line, marked by $2$, is the boundary of two different non-trivial phases, and this boundary follows the relation \[h_1 = \frac{\ln\left(2|-t \exp(h_2)-\Delta|\right)}{V}.\] These analytical expressions for the phase boundaries are obtained by substituting the hopping parameter $t = te^{-h_2}$ in the mathematical formulation proposed in Refs. \cite{PhysRevB.103.214202, PhysRevLett.110.176403}. 

Furthermore, the quasiperiodic potential, where the parameter $\beta$ is an irrational number, causes localization in the system. The system's localization is characterized by a generalized fractal dimension, which is determined in the following way. First, we partition the components of the eigenstates into $L_d$ number of boxes. Each box will then contain $d=L/L_d$ components. Consider the $i$-th eigenstate of the Hamiltonian $|\phi_i\rangle$, which can be written in the site basis $\{|n\rangle,\, n = 1, \dots, N\}$, as $|\phi_i\rangle = \sum_n c_{in} |n\rangle$. Here, $c_{in}$ represents the $n$-th component of the $i$-th eigenstate. Therefore, the probability associated with the $k$-th box of the $i$-th eigenstate is
\[p_k(d) = \sum\limits_{n = (k-1)d + 1}^{kd} |c_{in}|^2, ~{\rm where}~ k = 1, \dots, N_d.\]
Here, $|c_{in}|^2 = |u_{i,n}|^2 + |v_{i,n}|^2$ determines the occupation of the site $n$ for the $i$-th eigenstate, where $\{u_{i,n},\, v_{i,n}\}$ are the coefficients of the $i$-th eigenstate in the BdG basis. The generalized fractal dimension $D_q$ of any eigenstate is determined using the following relation:
\begin{equation}
D_q = \frac{1}{q-1}\lim_{d \rightarrow 0} \frac{\log\left(\sum\limits_{k=1}^{N_d} [p_k(d)]^q\right)}{\log d}.
\label{D_q_eq}
\end{equation}
In the finite-dimensional case, $D_q$ is determined from the slope of the numerator versus denominator curve given in Eq. \eqref{D_q_eq}. The fractal dimension $D_q \simeq 1.0 \, (\simeq 0)$ for the delocalized (localized) eigenstates. If $D_q$ is independent of $q$, then the corresponding eigenstate is a mono-fractal or simply fractal; otherwise, the eigenstate is multifractal. However, here, we study only the fractal nature of the eigenstates by calculating the fractal dimension $D_2$. In Fig. \ref{Phase}(b), we present the results of $\overline{D}_2$ as a function of the non-Hermitian parameters $h_1$ and $h_2$, where $\overline{D}_2$ is the average of $D_2$ over all the eigenstates. Here again, we observe three regions: delocalized, critical, and localized, based on the average fractal dimension $\overline{D}_2$. The delocalized region coincides with the topological trivial region with the winding number $w = 0$. The critical region is associated with the fractional winding number $w = 0.5$, while the localized region corresponds to $w = -1$.

Besides topological and localization transitions, we also observe unconventional real to complex transitions by calculating the energy eigenvalue with the largest imaginary part, and the corresponding result is presented in Fig. \ref{Phase}(c) \cite{PhysRevB.103.104203}. We observe real energy eigenvalues in the same region of $(h_1, h_2)$ parameters, where topologically trivial and delocalized regions were observed. In contrast, the eigenvalues become complex for topologically non-trivial and localized regions. Thus, we observe two regions in the non-Hermitian parameters space $(h_1, h_2)$ separated by a phase boundary, where the system is topologically trivial, delocalized, and energy eigenvalues are real-valued. The system is non-trivial and localized in the other regions with complex energy eigenvalues. The system makes triple-phase transitions through that phase boundary. 
However, as we increase the parameter $\Delta$, we do not observe the triple-phase transitions. We have presented the phase diagrams for $\Delta = 0.1$ and $0.5$ in Appendix \ref{AppendixA}.

\section{In-gap states due to asymmetric hopping}
\label{sec4}

\begin{figure*}
\begin{tabular}{cc}
    \includegraphics[width=0.69\linewidth]{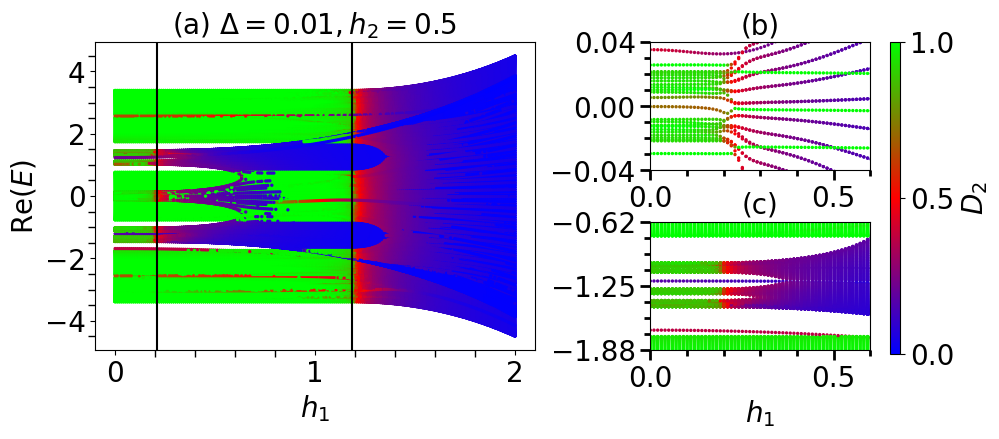}&
    \includegraphics[width=0.28\linewidth]{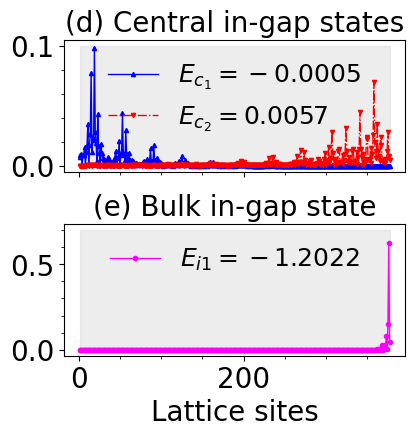}
\end{tabular}
\caption{Here, the non-Hermitian parameter $h_2 = 0.5$ and pairing strength $\Delta = 0.01$. (a) The real part of the energy as a function of the non-Hermitian parameter $h_1$ is presented, where the black dashed vertical lines indicate transition points. (b) An enlarged view of (a) is presented, focusing on the region around the central energy and revealing two central in-gap states. (c) Another zoomed-in view of (a) is presented, focusing on the region with bulk in-gap states, where a localized in-gap state is visible within a gap in the bulk. (d) The eigenstates of the central in-gap states are presented for $h_1 = 0.1$, where each state is localized at one of the ends of the system. (e) The eigenstate of one of the bulk in-gap states exhibits localization toward the right. Once again, we consider $h_1 = 0.1$. In (d) and (e), the energies of the central and bulk in-gap states are mentioned in the figure.}
\label{ES_OBC_1}
\end{figure*}

\begin{figure*}    
\begin{tabular}{cc}
    \includegraphics[width=0.69\linewidth]{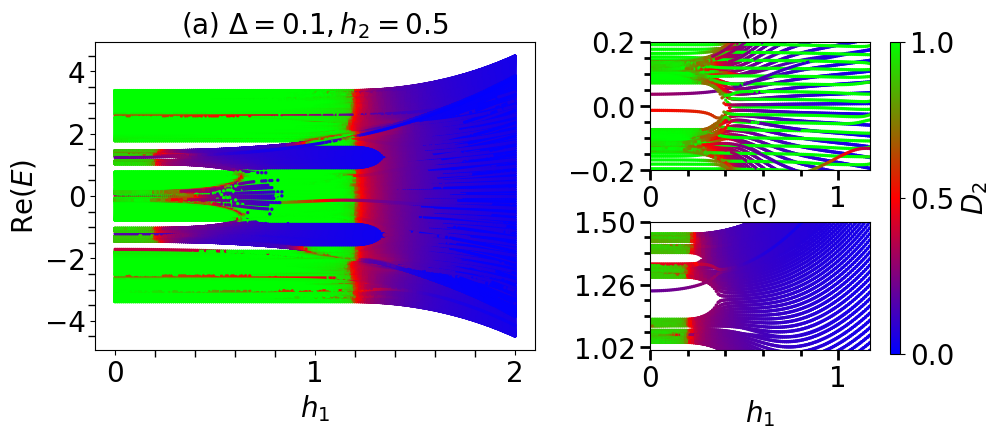}&
    \includegraphics[width=0.28\linewidth]{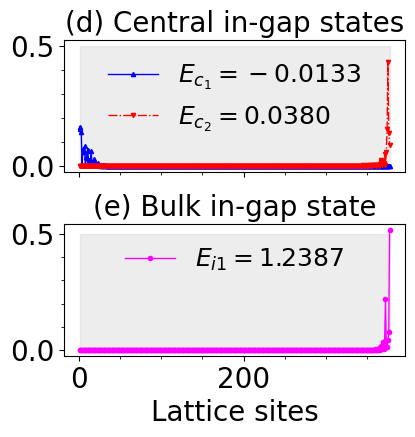}
\end{tabular}
\caption{We set the non-Hermitian parameter $h_2 = 0.5$, and consider stronger pairing strength $\Delta = 0.1$. Here, (a)-(e) present the same quantities as presented in Fig. \ref{ES_OBC_1}. In (d) and (e), the eigenvectors corresponding to the selected energy eigenvalues are mentioned in the figure.}
\label{ES_OBC_2} 
\end{figure*}

\begin{figure*}     
\begin{tabular}{cc}
    \includegraphics[width=0.69\linewidth]{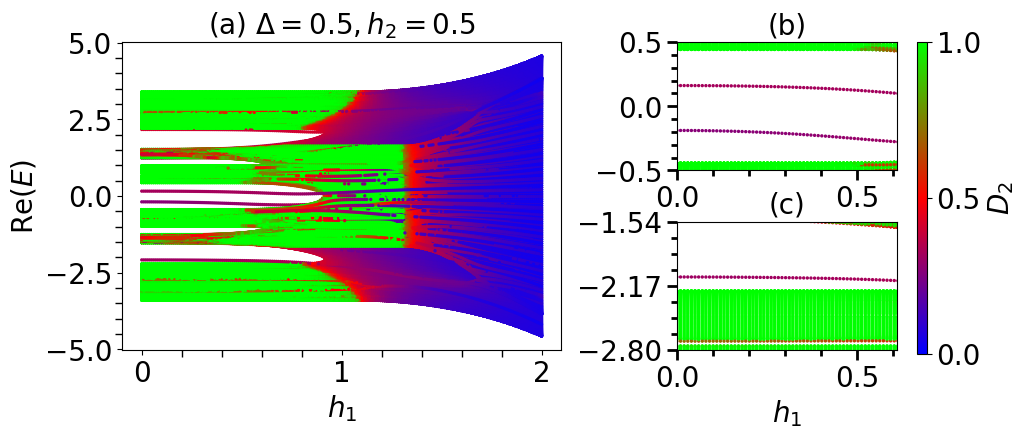}&
    \includegraphics[width=0.28\linewidth]{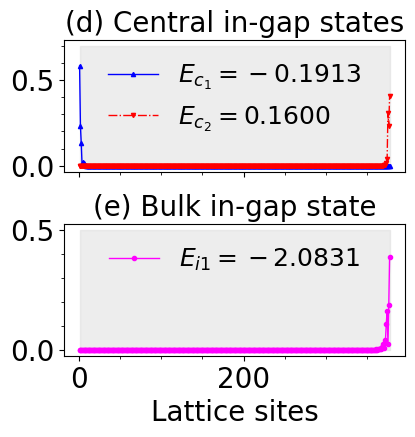}
\end{tabular}
\caption{Once again, the same quantities are presented in this figure as in the last two figures. We also keep the value of the non-Hermitian parameter the same at $h_2 = 0.5$ but consider a much stronger pairing strength $\Delta = 0.5$.}
\label{ES_OBC_3} 
\end{figure*}

\begin{figure*}
    \includegraphics[width=1\linewidth]{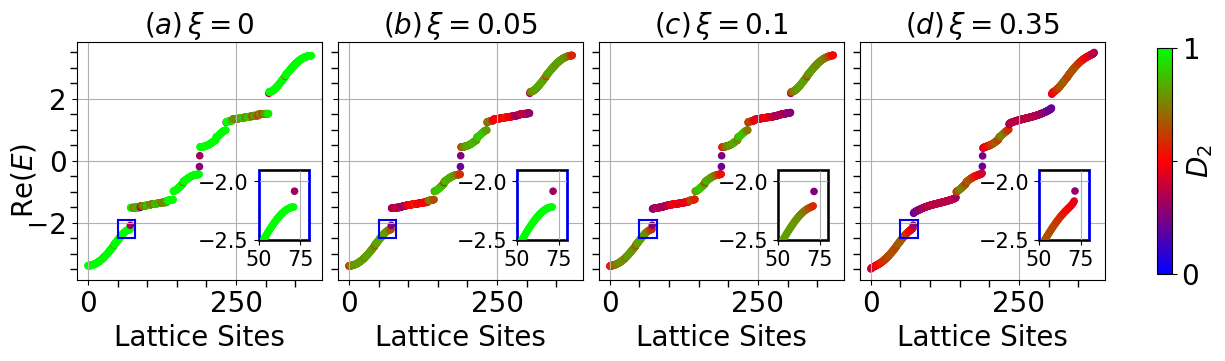}
    \caption{(a)-(d) exhibits the real part of the energy spectrum for different values of disorder strength. The inset shows an enlarged view of the bulk in-gap state.}
    \label{disorder}
\end{figure*}

This section studies the real part of energy eigenvalues of the Hamiltonian kernel $H^{\rm BdG}$ under OBC as a function of the non-Hermitian parameter $h_1$. This analysis highlights a striking transformation in the behavior of the edge states when the system is transitioning from symmetric to asymmetric hopping. For the case of symmetric hopping ($h_2=0$), the system is known to host robust MZMs \cite{PhysRevB.103.214202}. However, for the case of asymmetric hopping ($h_2 \neq 0$), the MZMs are no longer observed. Instead, the eigenspectrum reveals two distinct types of in-gap states: central and bulk in-gap states. As the nomenclature suggests, the central in-gap states are observed around the central gap, while the bulk in-gap states are observed within the gap in the bulk. The in-gap states were previously observed in the SSH-trimers model \cite{PhysRevLett.42.1698,PhysRevB.110.125424, Ghuneim_2024, PhysRevA.99.013833,PhysRevB.106.085109} and superconductors \cite{doi:10.1021/acs.nanolett.4c01581}.

Figure \ref{ES_OBC_1}(a) presents the real part of energy eigenvalues for a fixed hopping aymmetry $h_2=0.5$ and pairing strength $\Delta = 0.01$. The color scale represents the fractal dimension $D_2$, providing insights into the localization properties of the states. The black dashed lines correspond to the same transition lines shown in the phase diagram in Fig. \ref{Phase}. The presence of in-gap states is evident from the red or blue lines in the delocalized and critical region. Notably, these in-gap states are absent in the case of PBC, as discussed in Appendix \ref{AppendixB}. Since the pairing strength is weak, the superconducting gap is small. Therefore, to better visualize these in-gap states, Figs. \ref{ES_OBC_1}(b) and (c) present enlarged views of the spectrum near the central energy and bulk gaps, respectively. These zoomed-in plots reveal the presence of central and bulk in-gap states. For further investigation of the nature of these states, their wave functions are presented in Figs. \ref{ES_OBC_1}(d) and (e). The wave functions show that the central in-gap states are strongly localized, with one state at each end of the lattice. In contrast, the bulk in-gap state exhibits localization only at the right end of the system.

Similarly, Figs. \ref{ES_OBC_2} and \ref{ES_OBC_3} display the eigenspectra for stronger pairing strengths with $\Delta = 0.1$ and $0.5$, respectively. Here, we have not varied the non-Hermitian parameter and set it again at $h_2 = 0.5$. As the pairing strength increases, the superconducting gap widens, evident from the separation between the in-gap states. In addition, the central in-gap states become more localized at the lattice ends with increasing pairing strength, as shown in Figs. \ref{ES_OBC_2}(d) and \ref{ES_OBC_3}(d). Unlike Fig. \ref{ES_OBC_1}, the phase boundaries are not marked because our analytical formula can not capture the phase transition boundaries for stronger pairing strength. The corresponding PBC results are provided in Appendix \ref{AppendixB}, where these in-gap states are absent. In the following section, we investigate the robustness of these in-gap states, providing insights into their stability and topological nature.

\section{Effect of disorder on in-gap states}
\label{sec5}

A key feature of the topologically protected edge states is their resilience to disorder perturbation to the original Hamiltonian. In this section, we investigate the robustness of the central and bulk in-gap states in the presence of disorder. The disorder is introduced into hopping, potential, and pairing terms, ensuring its magnitude is smaller than the superconducting energy gap. Specifically, we introduce the random disorder factor $r$, where $r = 1 + \delta r$, and $\delta r$ is a uniformly distributed random variable within the range $[-\xi,\xi]$. Here, $\xi \in \mathbb{R}$ represents the disorder strength. Thus, the onsite potential $V$, hopping strength $t$ and the pairing strength $\Delta$ in the original Hamiltonian is replaced by $V_{\rm disordered} = rV$, $t_{\rm disordered}= rt$ and $\Delta_{\rm disordered} = r \Delta$. Here, we focus on the case of $\Delta = 0.5$ and $h_2=0.5$, considering disorder strengths $\xi = 0.05,\,0.1$ and $0.35$. These values are chosen so that the disorder strength remains within the range of the superconducting gap, which is approximately $0.4$.

Figure \ref{disorder}(a) shows the energy eigenvalues for the non-Hermitian parameter $h_1 = 0.1$ in the absence of disorder. In this case, we observe the central in-gap states and the bulk in-gap state, as previously shown in Fig. \ref{ES_OBC_3}(d) and (e). The bulk in-gap state from Fig. \ref{ES_OBC_3}(e) is marked with a blue rectangular box, and the inset offers a zoomed-in view to highlight its precise location within the bulk gap. These in-gap states provide the basis for studying how they behave when the disorder is introduced. In Fig. \ref{disorder}(b), we introduce a very weak disorder, with disorder strength $\xi = 0.05 $, and observe that the central and bulk in-gap states remain largely unaffected. The other eigenvalues also 
remain unchanged. As we increase the disorder strength to $\xi = 0.1 $, shown in Fig. \ref{disorder}(c), and then to $\xi = 0.35$, shown in Fig. \ref{disorder}(d), we observe that the central and bulk in-gap states remain robust, which is further demonstrating topological protection of these states. However, there are slight modifications in the bulk states as the disorder is increased. Specifically, the bulk states exhibit some deformation and tend to localize, indicated by color. Thus, we conclude that the central and bulk in-gap states observed in the eigenspectra are robust against disorder and are topologically protected.

\section{Conclusion}
\label{Conclusion}

In this work, we have explored the NHAAH model with unconventional superconductivity described by $p$-wave pairing, where non-Hermiticity is included in complex quasiperiodic potential and asymmetric hopping. We have particularly investigated the behavior of phases to understand the interplay between pairing and non-reciprocal hopping strengths. The system exhibits a triple phase transition at a small pairing strength $\Delta$, where topological, MI, and unconventional real-to-complex energy eigenvalues transitions coincide. Importantly, as we increase the pairing strength, the three transitions no longer coincide, and theoretical prediction fails to identify the phase boundary. However, for symmetric hopping with $h_2=0$, the analytical expressions for the phase boundaries remain valid even at higher pairing strengths. This underscores the asymmetric hopping as a key driver that disrupts the alignment between theoretical prediction and exact numerics. So, there is a competition between the asymmetric hopping and pairing strength in defining the phases. A striking feature of this system is the emergence of in-gap states in the presence of asymmetric hopping. Unlike the case of symmetric hopping with $h_2 = 0$, where robust MZMs appear, asymmetric hopping eliminates MZMs and gives rise to two distinct types of in-gap states: central in-gap states and bulk in-gap states. These in-gap states remain robust despite strong disorder in the system, highlighting their topological nature. Therefore, the observed in-gap states, being topologically protected and robust against perturbations, are promising candidates for topological quantum computational protocols. Unlike MZMs, which are challenging to realize experimentally due to strict symmetry requirements, these in-gap states offer a more flexible and accessible alternative.

\begin{acknowledgments}
JNB acknowledges financial support from DST-SERB, India, through a Core Research Grant CRG/2020/001701 and a MATRICS grant MTR/2022/000691. 
\end{acknowledgments}

\appendix

\section{Phase Diagrams for $\Delta = 0.1$ and $0.5$}
\label{AppendixA}

\begin{figure*}        
\begin{tabular}{ccc}
    \includegraphics[width=0.34\linewidth]{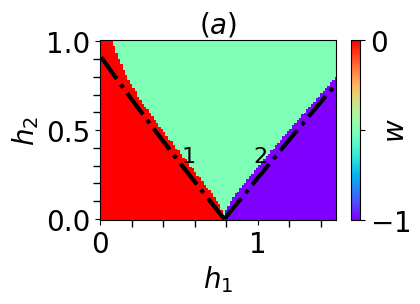}&
    \includegraphics[width=0.29\linewidth]{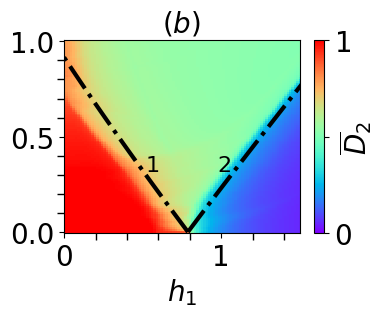}&
    \includegraphics[width=0.31\linewidth]{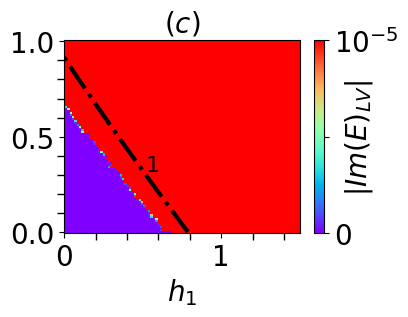}
\end{tabular}
\caption{The phase diagrams are presented for topological, MI, and unconventional real to complex transitions as a function of non-Hermitian parameters $h_1$ and $h_2$, for $\Delta = 0.1$. We observe that now all three transitions are not determined by transition lines $1$ and $2$.}
\label{Phase_b}
\end{figure*}

\begin{figure*}        
\begin{tabular}{ccc}
    \includegraphics[width=0.34\linewidth]{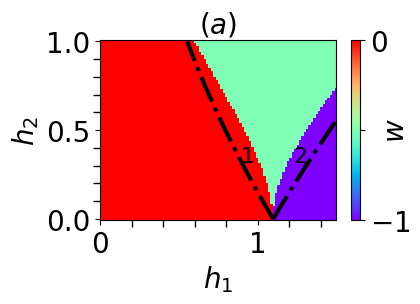}&
    \includegraphics[width=0.29\linewidth]{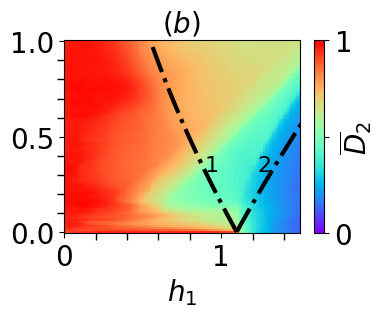}&
    \includegraphics[width=0.31\linewidth]{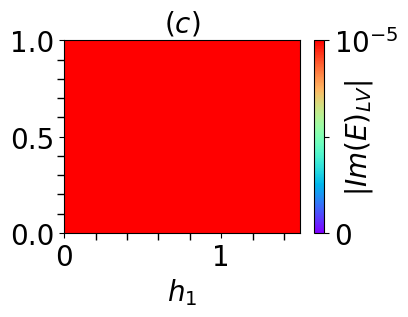}
\end{tabular}
\caption{The phase diagrams illustrate the topological, MI, and unconventional real to complex transitions as a function of the non-Hermitian parameters $h_1$ and $h_2$, for $\Delta = 0.5$. Notably, all three transitions deviate from the predicted transition lines. Furthermore, the entire region becomes complex due to the influence of strong pairing interactions.}
\label{Phase_c}
\end{figure*}

\begin{figure*}        
\begin{tabular}{ccc}
    \includegraphics[width=0.3\linewidth]{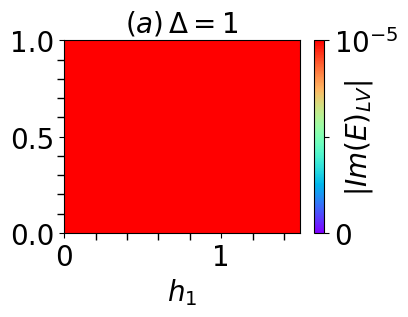}&
    \includegraphics[width=0.3\linewidth]{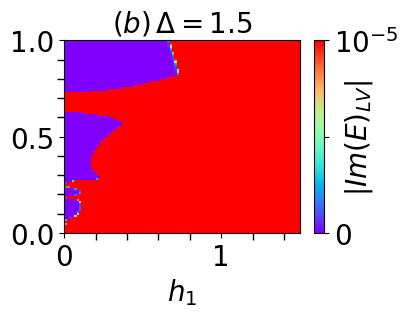}&
    \includegraphics[width=0.3\linewidth]{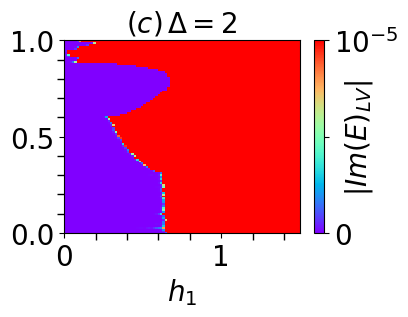}
\end{tabular}
\caption{The phase diagram highlights the unconventional real to complex transition in energy eigenvalues for $\Delta = 1.0,\,1.5$, and $2.0$. Notably, as the pairing strength increases, the eigenvalues revert to being real, demonstrating a re-entrant real energy region.}
\label{Real_complex}
\end{figure*}

In the main text, we presented a phase diagram illustrating a triple phase transition for $ \Delta = 0.01 $. We extend this analysis by presenting phase diagrams for stronger pairing strengths $\Delta = 0.1$ and $\Delta = 0.5$, as shown in Figs. \ref{Phase_b} and \ref{Phase_c}. For these cases, we observe that the transition lines $1$ and $2$, calculated analytically in the main text, no longer accurately describe the phase transitions. However, it is noteworthy that, for $h_2 = 0$, the phase diagrams corresponding to topological and MI transitions still match the analytical expression, indicating a double phase transition. As the pairing strength increases to $\Delta = 0.1$, the region with real energy eigenvalues shrinks, as shown in Fig. \ref{Phase_b}(c). For sufficiently strong pairing strength $\Delta = 0.5$, Fig. \ref{Phase_c} shows that all the energy eigenvalues become complex. However, as established in Ref. \cite{gandhi2024superconductingpwavepairingeffects}, the real energy region re-enters with further incrementing the pairing strength. To illustrate this re-entrant behavior, we present results for pairing strengths $\Delta = 1.0$, $\Delta = 1.5$, and $\Delta = 2.0$ in Fig. \ref{Real_complex}. The results show that the real eigenvalue region reappears in cases of $\Delta = 1.5$ and $\Delta = 2.0$.

\section{Eigenspectra analysis under PBC}
\label{AppendixB}

\begin{figure*}
\includegraphics[width=0.9\linewidth]{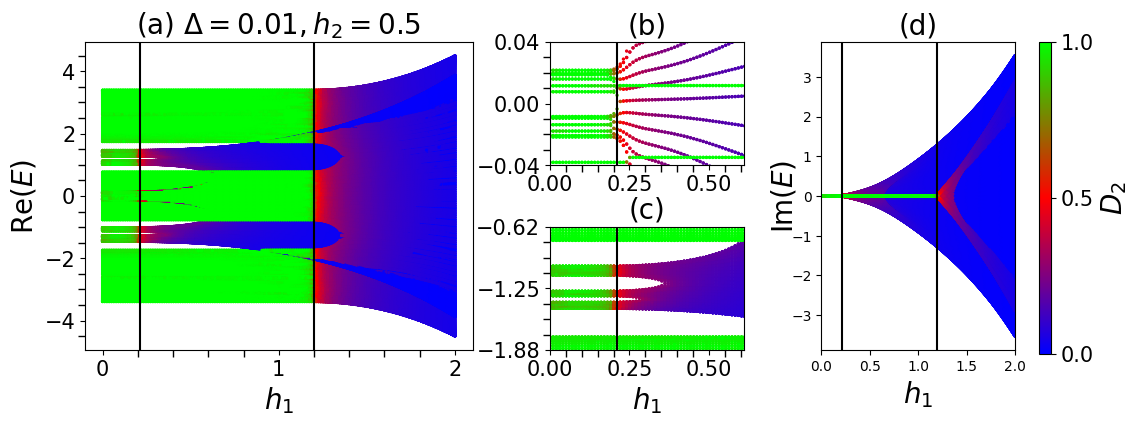}
\caption{The energy spectrum is studied for the non-Hermitian parameter $h_2 = 0.5$ and pairing strength $\Delta = 0.01$ under PBC conditions. (a) The real part of the energies as a function of the non-Hermitian parameter $h_1$ is presented. (b) and (c) Provide enlarged views of the spectrum, highlighting the absence of central energy in-gap states and bulk in-gap states. (d) The imaginary part of the energies is presented as a function of $h_1$.}
\label{ES_PBC_1}
\end{figure*}

\begin{figure*}
\includegraphics[width=0.9\linewidth]{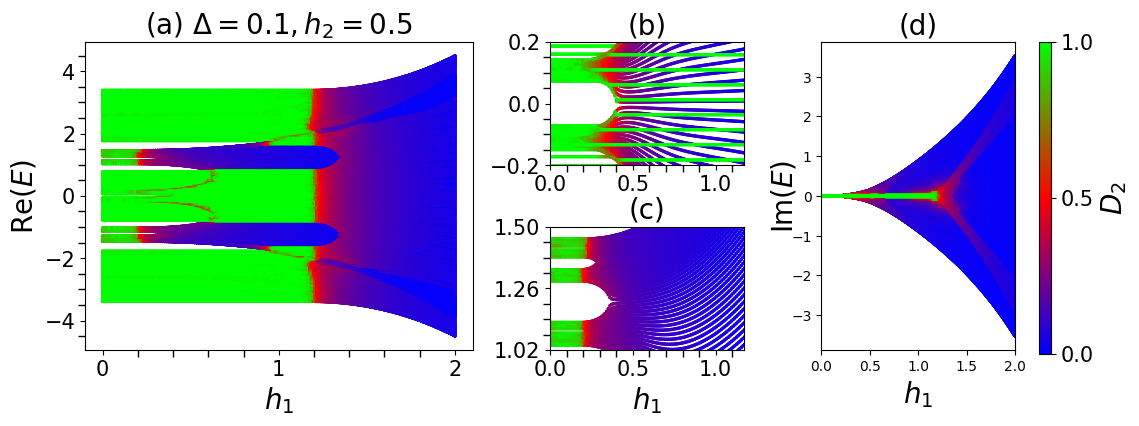}
\caption{The same results as in Fig. \ref{ES_PBC_1} are presented for stronger pairing strength $\Delta = 0.1$.}
\label{ES_PBC_2}
\end{figure*}

\begin{figure*}
\includegraphics[width=0.9\linewidth]{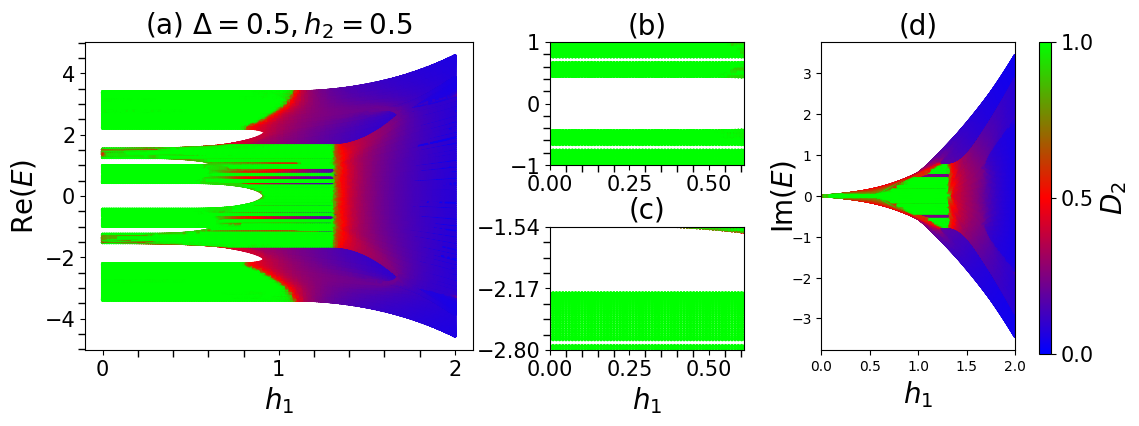}
\caption{Once again, the same results are presented as in the previous two figures for much stronger pairing strength $\Delta = 0.5$.}
\label{ES_PBC_3}

\end{figure*}

In the main text, Sec. \ref{sec4} discusses the behavior of the real part of the energy eigenvalues as a function of the non-Hermitian parameter $h_1$ under OBC. To differentiate the in-gap edge states from the bulk states, this Appendix provides the eigenspectra for the case of PBC. In Fig. \ref{ES_PBC_1}(a), the real part of the energy is shown as a function of $h_1$, for $\Delta = 0.01$, with the black dashed lines indicating the same transition points as discussed in the main text. The system resides in a delocalized phase in the region before the left vertical line. The intermediate region, between the two vertical lines, exhibits both delocalized and localized states separated by mobility edges. This region is referred to as the critical region. After the second vertical line, the system is predominantly in the localized phase. Figures \ref{ES_PBC_1}(b) and \ref{ES_PBC_1}(c) show enlarged views of the spectrum near the central energy and the bulk energy gap, which is analogous to the regions highlighted in Fig. \ref{ES_OBC_1}(b) and (c) for the OBC case. In comparison, it is evident that the central in-gap and bulk in-gap states observed under OBC are absent in the PBC case. Figure \ref{ES_OBC_1}(d) presents the imaginary part of the energy eigenvalues. The first vertical line marks the onset of a transition from real to complex energy eigenvalues. It is evident from the figure that the region with purely real eigenvalues corresponds to the delocalized phase. Figures \ref{ES_PBC_2} and \ref{ES_PBC_3} present the same result as the previous figure, but for stronger pairing strengths $\Delta = 0.1$ and $\Delta = 0.5$, respectively. Unlike the OBC case, as shown in Figs. \ref{ES_OBC_2} and \ref{ES_OBC_3}, in-gap states are absent in case of PBC. Furthermore, as illustrated in Fig. \ref{ES_PBC_3}(d), the energy spectrum of the system is completely complex for $\Delta = 0.5$. Hence, the absence of in-gap states under PBC demonstrates their edge-localized nature, which is consistent with the system's topological properties.

\bibliography{references}

\end{document}